\documentclass[aps,prl,twocolumn]{revtex4-1}
\usepackage{graphicx}
\usepackage{dcolumn}
\usepackage{color}
\usepackage{hyperref}

\begin{document}
\preprint{Draft version \today}

\title{Renyi Entropy of the Interacting Fermi Liquid}

\author{Jeremy McMinis}
\author{Norm M. Tubman}
\affiliation{Department of Physics, University of Illinois, Urbana IL}

\date{\today}

\begin{abstract}
We perform quantum Monte Carlo calculations to determine how the Renyi entropies, $S_n$, of the interacting Fermi liquid depend on Renyi order, $n$, and scale as a function of system size, $L$. Using the swap operator and an accurate Slater-Jastrow wave function, we compute Renyi entropies for spinless fermions interacting via the Coulomb and modified P\"{o}schl-Teller potentials across a range of correlation strengths. Our results show that interactions increase the Renyi entropies and increase the prefactor of their scaling laws. The relationships between Renyi entropies of different order $n$ are also modified. Additionally, we investigate the effect of the swap operator on the Fermi liquid wave function to determine the source of the $L\log L$ scaling form.
\end{abstract}

\maketitle
Entanglement is a central concept in quantum information theory, and has become increasingly important in studies of quantum many body models of condensed matter \cite{Neilson,mutual_info,op-2,prog1,tee-2,ee-pt,ee-spect}. 
The Renyi entropies computed from a spatially reduced density matrix provide a measure of entanglement, through non-local correlations in the ground state wave function \cite{mutual_info,op-2,note1}. The scaling laws for these entropies provide a ``fingerprint'' for quantum phases \cite{ee-pt,cc-eescale,log-term,scaling-tee,rmp-area}. For 1D systems these scaling laws can often be computed analytically using results from conformal field theory \cite{rmp-area,es-1d}. In higher dimensions the Renyi entropies are more difficult to compute and are often calculated using numerical techniques \cite{ dmrg,rmp-dmrg, peps,peps-ferm,kallin-1,ren-mol,QH-1,reyni-2,comp-2d,logterm-2d,spinless-2d,aflk-2d}. No general results for interacting systems in higher dimension are available, models must be addressed individually and results can differ greatly from their non-interacting counterparts \cite{kallin-1,ren-mol}. 

One of the most fundamental models in condensed matter is the Fermi liquid \cite{Pines}. The Renyi entropies of the non-interacting Fermi liquid in two dimensions have been conjectured to scale as $L \log L$ \cite{reyni-2,ee-proof,ee-proof2,widom-1,widom-2}. This scaling law is equivalent to the Widom conjecture and has been verified numerically for lattice and continuum Hamiltonians \cite{widom-1,widom-2,numfer2,widom-3,contferm}. High dimensional bosonization and conformal field theory studies have predicted that the scaling law of the Renyi entropies for the interacting and non-interacting Fermi liquid are identical \cite{widom-4,widom-5,highDbos}. These results imply that the relationship between $S_n$ of different order, $S_n=\frac{1}{2}(1+\frac{1}{n})S_1$, should also hold when interactions are included \cite{widom-4,es-1d}.

In this work, we use variational quantum Monte Carlo (VMC) to test these predictions. Specifically, we compute the Renyi entropy scaling forms for two different inter-particle potentials and compare them to the non-interacting results. VMC provides an accurate framework for the calculation of interacting and non-interacting quantum states in a many body framework, and is expected to produce reliable results for well optimized trial wave functions \cite{qmc-rmp}.  High quality trial wave function forms and robust optimization schemes have been developed to accurately compute observables \cite{qmc-rmp,rn-3d,dmc-sp-2d,opt}. These have been used to calculate benchmark energies for the electron gas and Fermi liquid parameters for the Coulomb potential, Helium-3, and several related Fermi liquid systems \cite{early-2d,rn-2d,qmc-rmp,rn-3d,fermiliquids}.

\textit{Hamiltonians and Trial Wave Functions-}
We choose two inter-particle potentials, the Coulomb and the modified P\"{o}schl-Teller potential (MPT), to test if our results are sensitive to the range of the interaction potential. Spin polarized electrons interact via the long range Coulomb potential \cite{Ashcroft}. The MPT potential is a well tested short range potential for the Fermi gas \cite{modpt,modpt2,modpt3}. If the range of the potential is important to the Renyi entropies then we could expect some qualitative difference between the scaling laws computed for these potentials.

The Hamiltonian for the spinless non-interacting electron gas in Hartree units is $H_{FG}=-\sum_{i}\frac{\nabla_i^2}{2}$. The exact fermionic ground state for this Hamiltonian is $\Psi_{FG}(R)=\langle R|(\prod_{k<k_f} c_k^\dagger|0\rangle)$. For the interacting electron gas $H_{HEG}=-\sum_{i}\frac{\nabla_i^2}{2} + \sum_{i<j}r_{ij}^{-1} + C(r_s)$, where $r_{ij}=|r_i-r_j|$ is the distance between electron $i$ and $j$, and the constant, $C(r_s)$, is due to a uniform positive background charge. For the MPT potential $H_{MPT}=-\sum_{i}\frac{\nabla_i^2}{2} + V_0\sum_{i<j} \cosh^{-2}(r_{ij})$. The parameter $V_0$ can be tuned to increase the correlation of the system and the effective range is set to the Wigner-Seitz radius, $r_s=1$ \cite{modpt}. 

We use a Slater-Jastrow trial wave function, $\mathcal{J}({R})\Psi_{FG}(R)$, for these interacting Hamiltonians. The Jastrow is of the form $\mathcal{J}({R}) = \textrm{exp} \left(\sum_{i<j} U(r_{ij})\right)$ and adds additional correlation between particle pairs. It is a reasonably accurate form that can be quickly computed \cite{qmc-rmp,fg2D-1,fg2D-2}. For the electron gas wave function, $\Psi_{HEG}({R})$, we eliminate the divergence in the electron-electron energy as $r_{ij}\rightarrow 0$ by including a cusp in the Jastrow \cite{rpa-1,cusp}. For the MPT wave function, $\Psi_{MPT}({R})$, there is no singularity in the potential and therefore no cusp in the Jastrow.

\textit{Renyi Entropy and the Swap Operator-}
The Renyi entropies are defined as
\begin{eqnarray}
S_n(\rho_A) = \frac{1}{1-n}\log \left(\textrm{Tr}(\rho_A^n)\right)
\end{eqnarray}
with the spatially reduced density matrix, $\rho_A$, of a system partitioned into two regions, $A$ and $B$. Using VMC we are able to compute unbiased estimates of $S_n$ for trial wave functions using the swap operator \cite{reyni-1},
\begin{eqnarray}
 \exp&&\left((1-n)S_n(\rho_A)\right) = \textrm{Tr}\left(\rho_A^n\right) \nonumber\\
 &&= \left\langle \Psi\otimes\Psi\cdots\otimes\Psi | \widehat{\textrm{ SWAP}_{n}} |\Psi\otimes\Psi\cdots\otimes\Psi \right\rangle.
\end{eqnarray}
The swap operator computes $\textrm{Tr}(\rho_A^n)$ by cyclically swapping coordinates between $n$ statistically independent copies of the system, each sampling $\Psi^2$. For instance, we can compute $S_2$ using,
\begin{eqnarray}
 \textrm{Tr}(\rho_A^2)&=&\left\langle \Psi\otimes\Psi | \widehat{\textrm{ SWAP}_{2}} |\Psi\otimes\Psi \right\rangle \nonumber\\
 &=&\frac{ \int |\Psi(R^1)|^2|\Psi(R^2)|^2 \frac{\Psi(R_A^1,R_B^2)}{\Psi(R^1)}\frac{\Psi(R_A^2,R_B^1)}{\Psi(R^2)} } {\int |\Psi(R^1)|^2|\Psi(R^2)|^2}
\end{eqnarray}
where $R_A^i$ represents the electronic spin and spatial coordinates in region $A$ for copy $i$, $R^i=(R_A^i,R_B^i)$ and $\Psi(R)=\langle R | \Psi \rangle$. 

\textit{Computing Renyi Entropies-}
In practice, the expectation value of $\textrm{Tr}(\rho_A^n)$ decays rapidly as region $A$ increases in size \cite{reyni-2}. To compute Renyi entropies for large regions we factorize the swap operator into two parts: sign, $S_\sigma$, and magnitude, $S_M$ \cite{reyni-2}. These two pieces are computed separately then summed to obtain the full value. This factorization is not an approximation and does not change the values we compute for $S_n$. The magnitude component can be further factorized into $S_M=S_V+S_N$, with $S_N$ due to swapped copies having different numbers of particles in region $B$, and $S_V$ from the absolute value of the swap when they do have the same number. The estimators for the full factorization,
\begin{eqnarray}
 \frac{1}{1-n}\log\left(\left\langle \widehat{\textrm{ SWAP}_n} \right\rangle \right) = \left\langle S_N \right\rangle  + \left\langle  S_V \right\rangle +  \left\langle  S_\sigma \right\rangle\label{factored_swap},
\end{eqnarray}
can be found in the supplementary material.

\begin{figure}
\includegraphics[scale=0.7]{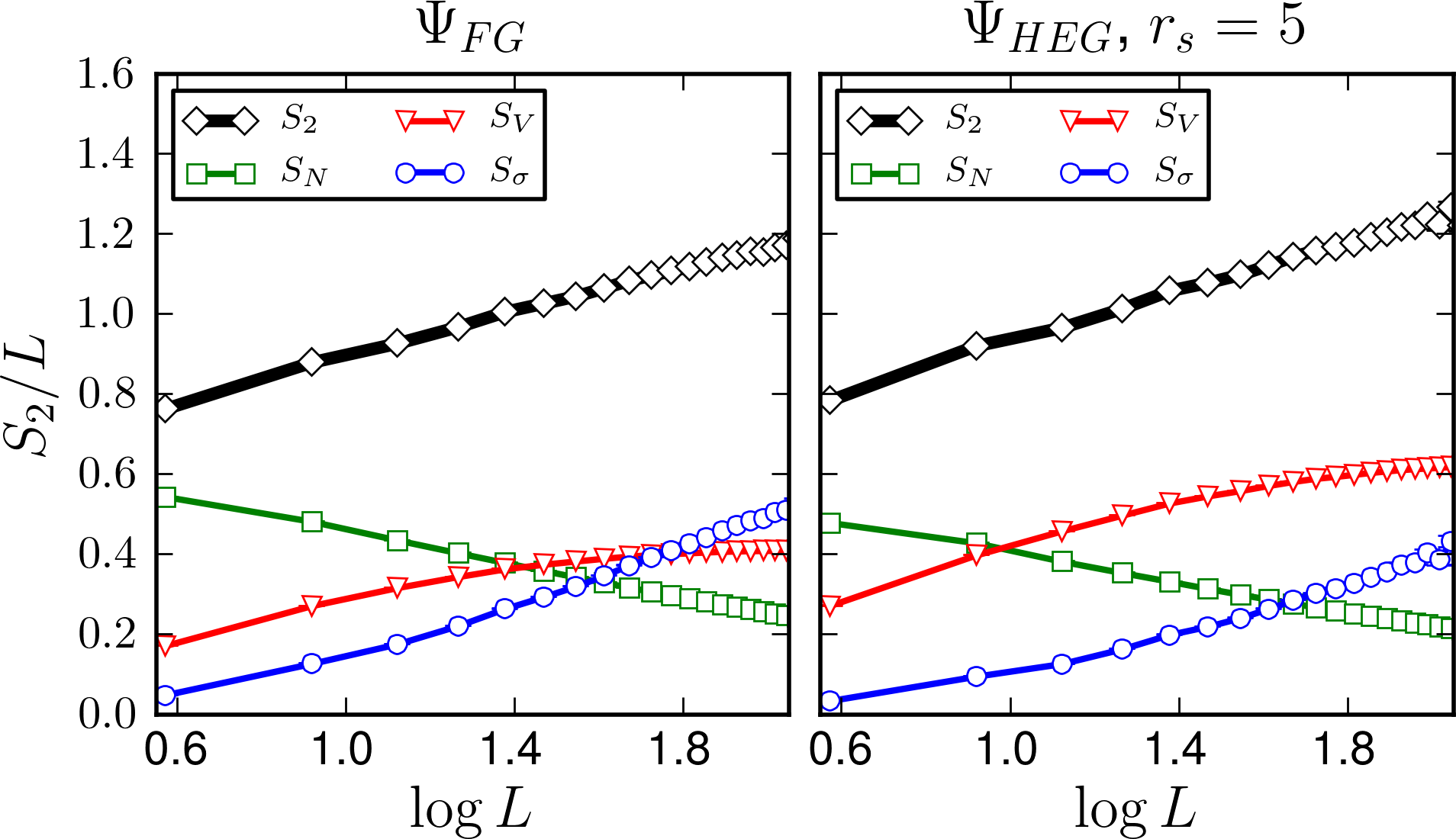}
\caption{  $S_2/L$ as a function of $\log L$, $L=\sqrt{\pi \langle N \rangle}$, for (\textit{Left}) the non-interacting electron gas and (\textit{Right}) the interacting gas at $r_s=5$. The bold black line is $S_2$. $S_N$, $S_\sigma$, and $S_V$ are the factorization of $S_2$ as in eqn. \ref{factored_swap}. The $\log L$ boundary law violation comes from the sign of the swap, it is the only term that continues to increase as $L$ gets large. For the interacting system at $r_s=5$, the Jastrow increases the contribution from $S_V$. \label{spo0_orb}}
\end{figure}

\begin{figure}
\includegraphics[scale=1.0]{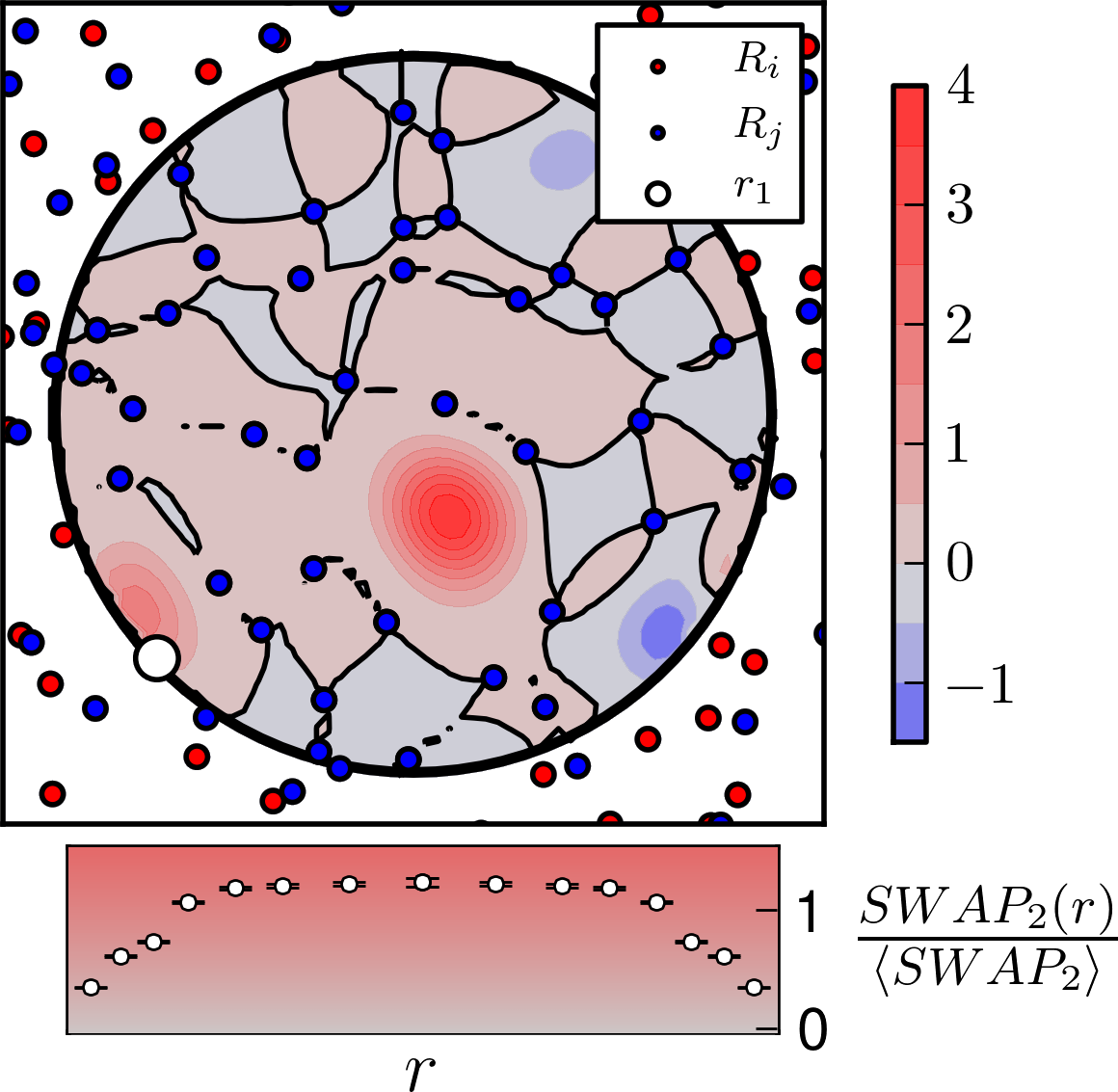}
\caption{(\textit{Top}) We scan a single particle from $r_1$ through the circular region $A$ and plot $\textrm{SWAP}_2(r_1)$, defined in the Letter body, for $\Psi_{FG}$. The white dot is the scanned particle at it's original position, $r_1$, in the lower left of Region $A$. The swap operator exchanges the red (blue) particles from copy $R^i$ ($R^j$) outside of region $A$. The red (blue) regions are positive (negative), with saturation representing magnitude. (\textit{Bottom}) The average amplitude of $\textrm{SWAP}_2(r)$, where $r$ is the distance from the center of region $A$. $\textrm{SWAP}_2(r)$ is larger than it's average value in the interior and decays rapidly near the boundary. The sign behaves similarly.
\label{swap_nodes}}
\end{figure}

\textit{Source of the Area Law Violation-}
In Fig. \ref{spo0_orb} we plot the contributions to $S_2$ from the factorization in eqn. \ref{factored_swap} for the non-interacting Fermi liquid and the interacting electron gas at $r_s=5$. The contribution to $S_2$ from the sign of the swap operator grows as $L$ increases, while the other contributions plateau or decrease. This indicates that as $L$ gets large the sign of the swap operator is the only component that can be responsible for the leading scaling behavior.  We note that previous studies found the same qualitative behavior \cite{reyni-2}. 

We investigate the spatial structure of the amplitude of the swap operator in Fig. \ref{swap_nodes}. The upper plot shows $\widehat{\textrm{ SWAP}_2}(r_1)=\Psi(R_1)\Psi(R_A^1,R_B^2)\Psi(R_2)\Psi(R_A^2,R_B^1)$, with $r_1$ a coordinate in $R_1$: $R_1=\{r_1,r_2,\ldots,r_{N_e}\}$. This plot is produced by swapping all particles in region $B$ then scanning a single particle initially located at $r_1$, through region $A$ and plotting the amplitude of the swap operator. The lower plot shows the normalized average amplitude of the swap operator as a function of distance from the center of region $A$. This average amplitude, just as the average sign, is approximately constant in the center and decays as particles are moved out to the edge. This suggests that the change in sign of the swap operator due to particles near the boundary drives the $\log$ violation of the area law. 

Heuristically, we can justify the Jastrow modifying the $L\log L$ and $L$ scaling terms.
The Jastrow can change the leading scaling law of the Renyi entropies by changing the particle distribution.
It can also create additional $L$ dependence by destroying the short range correlations in the Jastrow for the particles in region $A$ near region $B$.
We expect the difference in the Renyi entropies between the interacting and non-interacting systems to be well fit by these corrections, $\Delta {S_2}/L=(S_2^{HEG}-S_2^{FG})/L=l+m\log L$.

\textit{Computational Details-}
All QMC calculations are performed in QMCPACK\cite{qmcpack}. We parameterize the Jastrow by a flexible B-Spline function with fixed cusp conditions depending on the potential and optimize them using a variant of the linear method \cite{spline,opt}. A more accurate Slater-Jastrow-Backflow wave function was also tested \cite{bflow}. Using this wave function Renyi entropy results for $S_2$ at $r_s=5$ were the same as the Slater-Jastrow values. The only changes were in the orbital contributions to the Renyi entropy. Backflow increased the determinant and decreased the Jastrow contributions to the swap magnitude.

For the electron gas, the swap operator for $S_2$ was used without factorization for circular regions $A$ up to $\langle N\rangle =16$ for a simulation cell of $137$ electrons and with the sign factorization technique for $\langle N\rangle =16$, $25$, and $36$ in $261$ electron systems. We computed $S_2$ for regions of $\langle N\rangle =1$ though $16$ for the $137$ particle simulation cell using the unfactorized estimator for the MPT potential. When factorization is unnecessary we are able to increase our computational efficiency by computing the swap operator for several regions sizes at the same time. Approximately $10^6$ CPU hours were spent on the smaller systems and the same on the larger. For the higher order Renyi entropies we were restricted to $\langle N \rangle=16$ or fewer electrons for the $261$ electron system.

We rescale all lengths, $L\rightarrow L/(\sqrt\pi r_s)=\sqrt{\pi \langle N \rangle}$, so that $S_n(L)$ for the non-interacting Fermi liquid has no $r_s$ dependence. Under this rescaling, all interacting $S_n$ will collapse to the non-interacting $S_n$ if interactions are irrelevant. We use a $\log L$ x-axis to highlight the log linear nature of $S_n/L$ and $\Delta {S_n}/L$.

\textit{Results-}
We find that interactions increase the Renyi entropies and increase the prefactor in their scaling law. More strongly correlated Hamiltonians produce ground state wave functions with stronger particle correlations and larger Renyi entropies. We also find that Coulomb interactions change the simple relationship between the Renyi entropies of different order $n$.

The non-interacting data is well fit by the leading scaling form predicted by the Widom conjecture, $S(L)=m_0(L/l_0)\log(L/l_0)$ with $m_0=0.032(2)$ and $l_0=0.113(5)$ \cite{widom-1,widom-2,contferm}.
 
\begin{figure}
\includegraphics[scale=0.7]{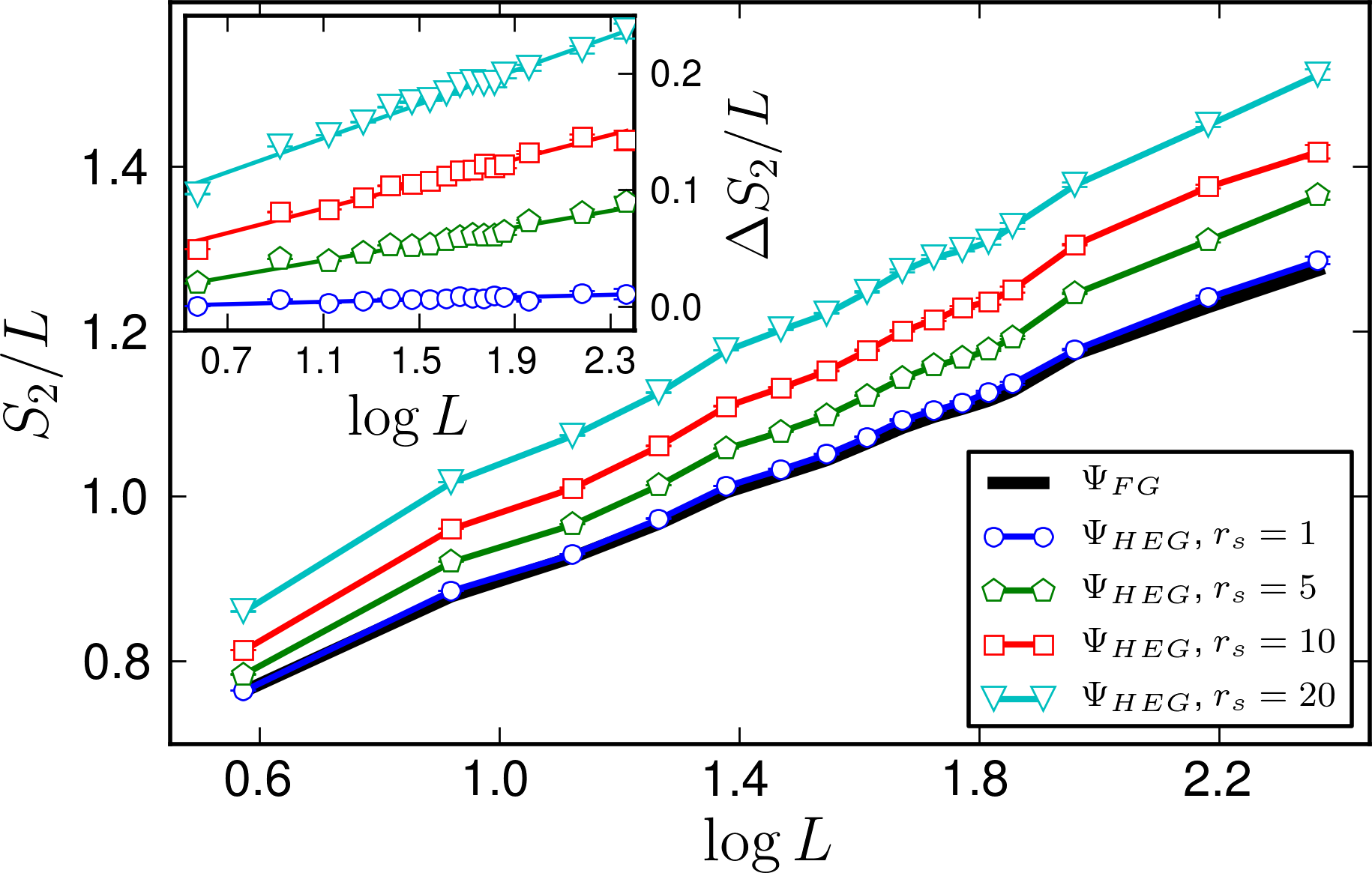}
\caption{(\textit{Main Plot}) $S_2/L$ for the spin polarized interacting electron gas at $r_s=1,5,10,20$ and the non-interacting Fermi liquid. $L$ is scaled so that without interactions all lines lie on top of $\Psi_{FG}$ results.  The high density $r_s=1$ data falls on top of the non-interacting liquid while the lower density, more strongly correlated, gasses have larger Renyi entropies. The $\Psi_{FG}$ reference is the exact non-interacting $S_2$ computed using the same number of particles as the interacting system for each $L$. (\textit{Inset}) $\Delta {S_2}/L=(S_2^{HEG}-S_2^{FG})/L$ plotted against $\log L$. The Renyi entropy for the interacting case appears to scale the same as the non-interacting case with a larger prefactor. The lines are from a fit to $\Delta {S_2}/L=l +m\log L$ shown in Table \ref{Cshifts}.
\label{s2_plot}}
\end{figure}

As shown in Fig. \ref{s2_plot}, Coulomb interactions modify the Renyi entropies when the Coulomb energy becomes larger in magnitude that the kinetic energy. 
At  $r_s=1$ the Coulomb energy is approximately equal to the kinetic energy and interactions do not significantly alter $S_2$.
As Coulomb interactions become stronger, the Renyi entropy $S_2$ increases.
For $r_s=20$ the Coulomb energy is an order of magnitude larger than the kinetic energy, and the Renyi entropy is increased relative to the non-interacting Fermi gas.
The higher order Renyi entropies, reported in the supplementary material and shown in Fig. \ref{sn_plot}, show the same qualitative trend.

\begin{figure}
\centering
\includegraphics[scale=0.7]{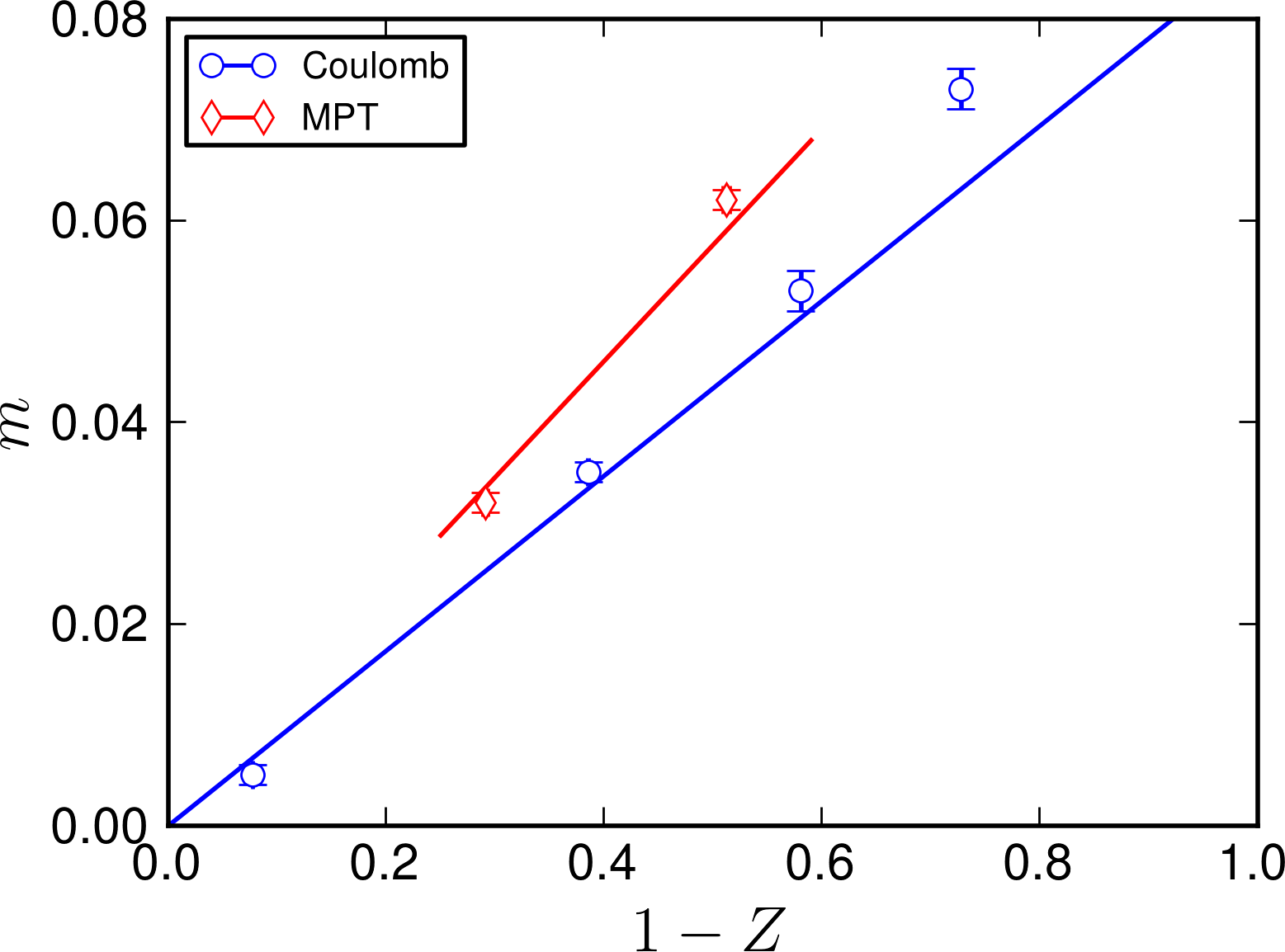}
\caption{Slope difference, $m$ ($\Delta S_2/L=l+m\log L$), as a function of $1-Z$, one minus the quasi particle renormalization factor, for the Coulomb and modified P\"{o}schl-Teller (MPT) potential at all correlation strengths considered. Lines are guides to the eye from linear fits constrained to go through the origin.
\label{Z_m}}
\end{figure}

We find $\Delta {S_2}/L$ is well fit to the two parameter form $l+m\log L$. The fitting parameters for the Coulomb potential at all $r_s$ are presented in Table \ref{Cshifts} and for the MPT at all $V_0$ in Table \ref{PTshifts}. The Coulomb fit is plotted in the inset of Fig. \ref{s2_plot}. As the strength of the inter-particle potential is increased so too is the prefactor of the leading scaling term, $m$. We compute the quasi particle renormalization factor, $Z$, to determine how strong inter-particle correlations are and compare scaling laws between different model potentials \cite{fg2D-2}. The scaling laws for the MPT potentials show qualitative agreement with the interacting electron gas with the same $Z$. As $Z$ increases so does the slope of $\Delta S_2/L$. These slopes are almost linear with $Z$ and are plotted in Fig. \ref{Z_m}. We note that finite size corrections to $Z$ may not preserve this relationship \cite{fg2D-2}.

On the left of Fig. \ref{sn_plot}, $\Delta S_n/L$ is plotted for the Coulomb potential for a region of size $\langle N \rangle = 6.25$. The simple relationship between non-interacting $S_n$ does not hold when Coulomb interactions are included. Interactions increase the Renyi entropies for all $n$. As shown on the right of Fig. \ref{sn_plot}, upon rescaling, $\frac{S_n}{L}\rightarrow\frac{2n}{1+n}\frac{S_n}{L}$, the non interacting Renyi entropy is made constant. The difference between the interacting and non-interacting $S_n$ decreases as $n$ gets larger. The growth of the error bars as $L$ and $n$ increase prevent a good fit for their scaling laws and the relationship between $S_n$ of different order $n$.

\begin{figure}
\includegraphics[scale=0.7]{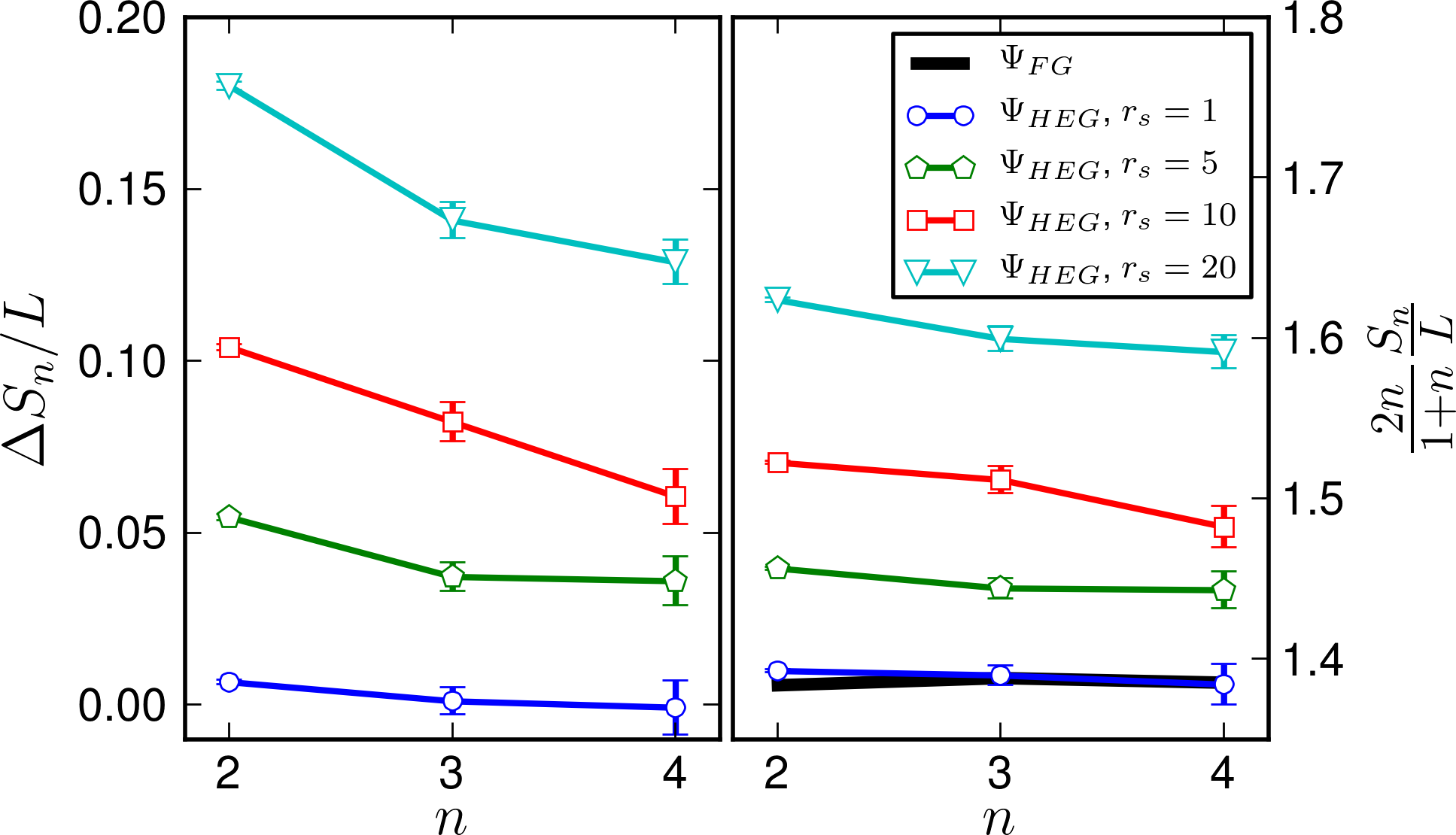}
\caption{For the Coulomb potential: (\textit{Left}) $\Delta S_n/L$, for $n=2,3,4$ for region $A$ size $\log L\approx1.5$, $\langle N\rangle =6.25$. $S_n$ increases as correlations increase and decreases as $n$ increases. (\textit{Right}) $\frac{2n}{1+n}\frac{S_n}{L}$ for $n=2,3,4$. Besides some small $L$ oscillations, this scaling makes the non-interacting Renyi entropies linear in $n$. Interactions modify this relationship and the rescaled Renyi entropies decrease as $n$ gets larger.
\label{sn_plot}}
\end{figure}

\begin{table}
\title{Coulomb Potential}
\begin{tabular}{|c|c|c|c|c|c|}
    \hline 
    $r_s$ & $m$ & $l$ & $ Z $ & Corr \\ \hline
    1  & 0.005(1) & 0.001(1) & 0.922(1) & 0.991 \\ 
    5  & 0.035(1) & 0.001(3) & 0.614(1) & 0.990 \\ 
    10 & 0.053(2) & 0.027(3) & 0.419(1) & 0.988 \\ 
    20 & 0.073(2) & 0.065(4) & 0.272(1) & 0.985 \\ \hline
\end{tabular}
\caption{$r_s$ is the Wigner-Seitz radius of the electron gas. $r_s=0$ is the non-interacting system. $\Delta {S_2}/L=l +m \log L$. $Z$ is the quasi particle renormalization factor. Corr$=(E_{VMC}-E_{HF})/(E_{DMC}-E_{HF})$ is the ratio of correlation energy the wave function recovers and provides a metric for wave function quality.\label{Cshifts}}
\end{table}

\begin{table}
\title{Modified P\"{o}schl-Teller Potential}
\begin{tabular}{|c|c|c|c|c|c|}
    \hline 
    $V_0$ & $m$ & $l$ &  $ Z $ & Corr \\ \hline
    10 & 0.032(1) & 0.007(2) & 0.708(3) & 0.984 \\ 
    20 & 0.062(1) & 0.028(1) & 0.487(4) & 0.981 \\
    \hline
\end{tabular}
\caption{$V_0$ is the strength of the potential for the modified P\"{o}schl-Teller (MPT) potential. All other labels are as in Table \ref{Cshifts}.\label{PTshifts}}
\end{table}

\textit{Error analysis-} For these calculations there are two primary sources of error, finite size effects, and trial wave function bias. Finite size effects come from two places, one is a function of the number of electrons in the system, the other is from subleading corrections to the Renyi Entropy scaling law. By integrating out regions that are a small fraction of the system size, we minimize interactions of region $A$ and it's periodic neighbors. 
For small subregions we see oscillations, small relative to $S_2$, that decay as the subregion length increases for both the interacting and non-interacting cases \cite{contferm}. Fits are performed on $\Delta S_2/L$ to minimize the effect of oscillations on the fitted parameters. 

The degree to which the trial wave function represents the Fermi liquid depends on it's overlap with the exact ground state wave function. We estimate the quality of our trial wave function by looking at the amount of correlation energy the VMC recovers relative to the fixed node diffusion quantum Monte Carlo result, $E_{corr}=E_{DMC}-E_{HF}$ \cite{qmc-rmp}. We recover more than $98\%$ of this correlation energy for all densities and potentials, shown in Table \ref{Cshifts} and \ref{PTshifts}. Also because the Slater-Jastrow and Slater-Jastrow-Backflow Renyi entropies are similar, we expect them to be accurate.

\textit{Conclusions and Future Work-}
In this work we have performed the first \textit{ab initio} calculation of the Renyi entropy of the interacting Fermi liquid. Our results show that, for this system, interactions modify the scaling behavior of the Renyi entropies. As particle correlations, quantified by $Z$, are strengthened, the prefactor for the leading scaling behavior of the Renyi entropies increases. These results hold for all wave function forms, correlation strengths, and inter particle potentials considered in this work. We also find the relationship between $S_n$ of different order $n$ changed. The source of the disagreement with recent theoretical predictions is not clear and deserves further investigation \cite{highDbos,widom-4,widom-5}. Work is currently underway improving the efficiency of computing higher order Renyi entropies and investigating related systems such as the paramagnetic electron gas and the Wigner crystal.

\textit{Acknowledgements-}
We would like to thank David Ceperley, Eduardo Fradkin, Sarang Gopalakrishnan, and Benjamin Hsu for conversations. This work was supported by the National Science Foundation and EFRC—Center for Defect Physics sponsored by the U.S. DOE, Office of Basic Energy Sciences. This work used the Extreme Science and Engineering Discovery Environment (XSEDE), which is supported by National Science Foundation grant number OCI-1053575., and resources provided by the Innovative and Novel Computational Impact on Theory and Experiment (INCITE) at the Oak Ridge Leadership Computing Facility at the Oak Ridge National Laboratory, which is supported by the Office of Science of the U.S. Department of Energy under Contract No. DE-AC05-00OR22725.


\begin{thebibliography}{55}
\bibitem [{not()}]{note1}%
  \BibitemOpen
  \href@noop {} {}\bibinfo {note} {All the information in the entanglement
  spectrum is also contained in the full set of Renyi entropies.}\BibitemShut
  {Stop}%
\end{thebibliography}
%

\onecolumngrid 
\section{Supplementary Material}
\subsection{VMC Estimator}
The VMC estimator can be factorized as,
\begin{eqnarray}
 S_2  &=& -\log\left(\left\langle \Psi^2\otimes\Psi^2\left| \widehat{\textrm{SWAP}_2}\right |\Psi^2\otimes\Psi^2\right\rangle \right)\\
  &=& -\log\left(\left\langle \widehat{\textrm{SWAP}_2} \right\rangle \right)\\
 &=& -\log \left(\left\langle  \delta_{N_1,N_2} \right\rangle \times  \frac{\left\langle|\widehat{\textrm{SWAP}_2}|\right\rangle}{\left\langle \delta_{N_1,N_2}\right\rangle}      \times  \frac{\left\langle\widehat{\textrm{SWAP}_2}\right\rangle}{\left\langle|\widehat{\textrm{SWAP}_2}|\right\rangle} \right)\\
&=& \left\langle S_N \right\rangle +\left\langle S_\sigma \right\rangle+ \left\langle S_V \right\rangle \label{factored_swap_supp}.\\
 \left\langle S_N \right\rangle &=& -\log \left(\left\langle \delta_{N_1,N_2} \right\rangle\right) \\
 \left\langle S_V \right\rangle &=& -\log \left( \frac{ \left\langle  |\widehat{\textrm{SWAP}_2}| \right\rangle }{ \left\langle \delta_{N_1,N_2} \right\rangle} \right) \\
 \left\langle S_\sigma \right\rangle &=& -\log \left(   \frac{ \left\langle\widehat{\textrm{SWAP}_2} \right\rangle }{ \left\langle |\widehat{\textrm{SWAP}_2}|\right\rangle  } \right)
\end{eqnarray}
where $S_N$ is the number fluctuation contribution, $S_\sigma$ is the sign contribution, and $S_V$ is the magnitude contribution. The Kronecker delta, $\delta_{N_1,N_2}$, is one when copy $1$ and $2$ have the same number of particles in region $B$ and zero otherwise. \\
To more efficiently compute the sign we sample a different probability distribution, the swap magnitude,
\begin{eqnarray}
 \left\langle S_\sigma \right\rangle &=& \frac{ \int |\Psi(R^1)|^2|\Psi(R^2)|^2 \left( \frac{\Psi(R_A^1,R_B^2)}{\Psi(R^1)}\frac{\Psi(R_A^2,R_B^1)}{\Psi(R^2)} / \left|\frac{\Psi(R_A^1,R_B^2)}{\Psi(R^1)}\frac{\Psi(R_A^2,R_B^1)}{\Psi(R^2)}\right|\right) } { \int |\Psi(R^1)|^2|\Psi(R^2)|^2  \left|\frac{\Psi(R_A^1,R_B^2)}{\Psi(R^1)}\frac{\Psi(R_A^2,R_B^1)}{\Psi(R^2)}\right| }
\end{eqnarray}
where $R_A^i$ represents the electronic spin and spatial coordinates in region $A$ for copy $i$, $R^i=(R_A^i,R_B^i)$ and $\Psi(R)=\langle R | \Psi \rangle$. This computes the sign of the swap operator from the regions in $R_1$, $R_2$ that contribute most to $S_2$. The magnitude and particle number contributions decay more slowly than the sign contribution and are sampled over the original distribution, $\Psi^2\otimes\Psi^2$.\\

\subsection{Higher Order Renyi}
\begin{figure}
\includegraphics[scale=0.8]{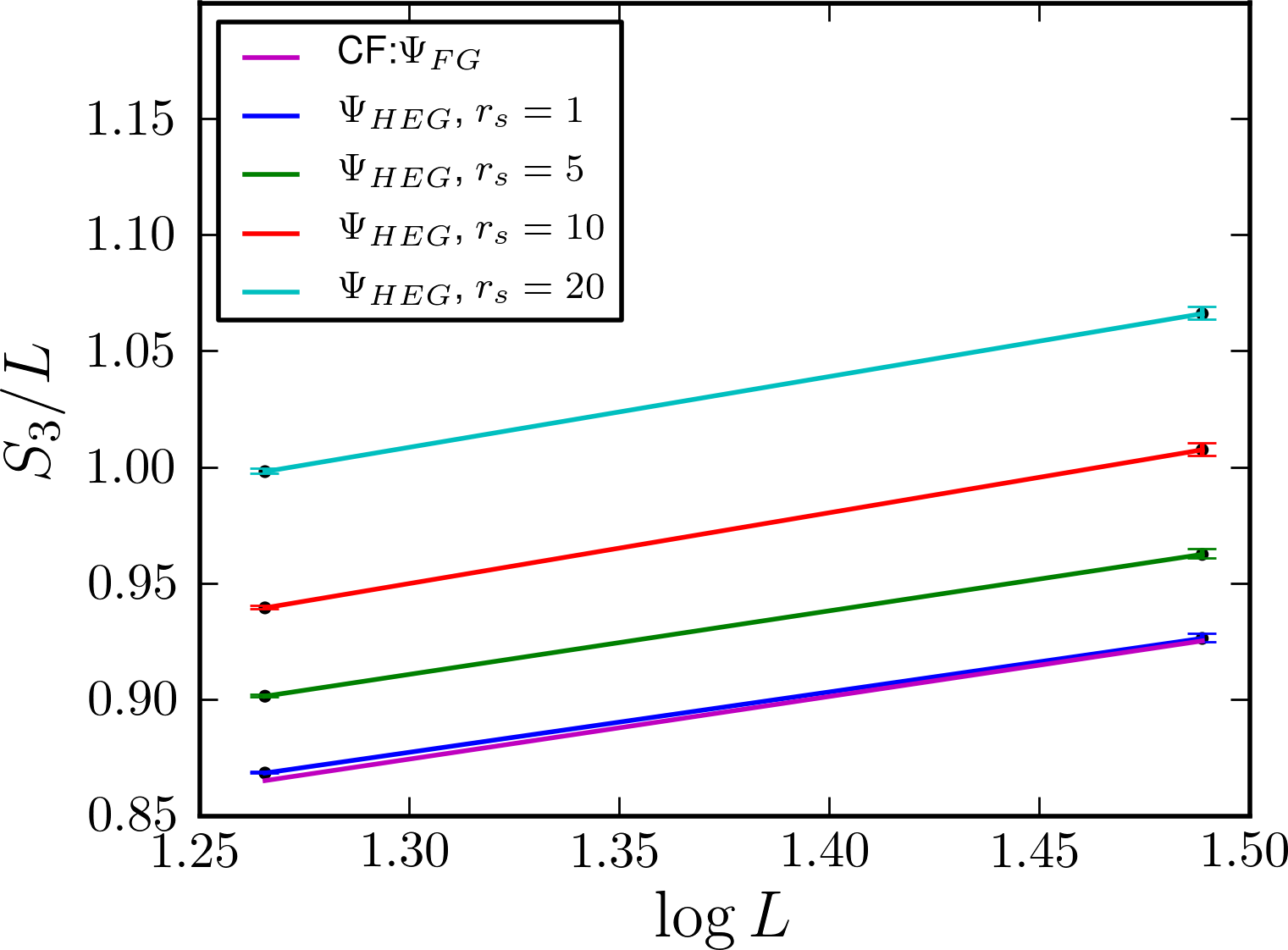}
\includegraphics[scale=0.8]{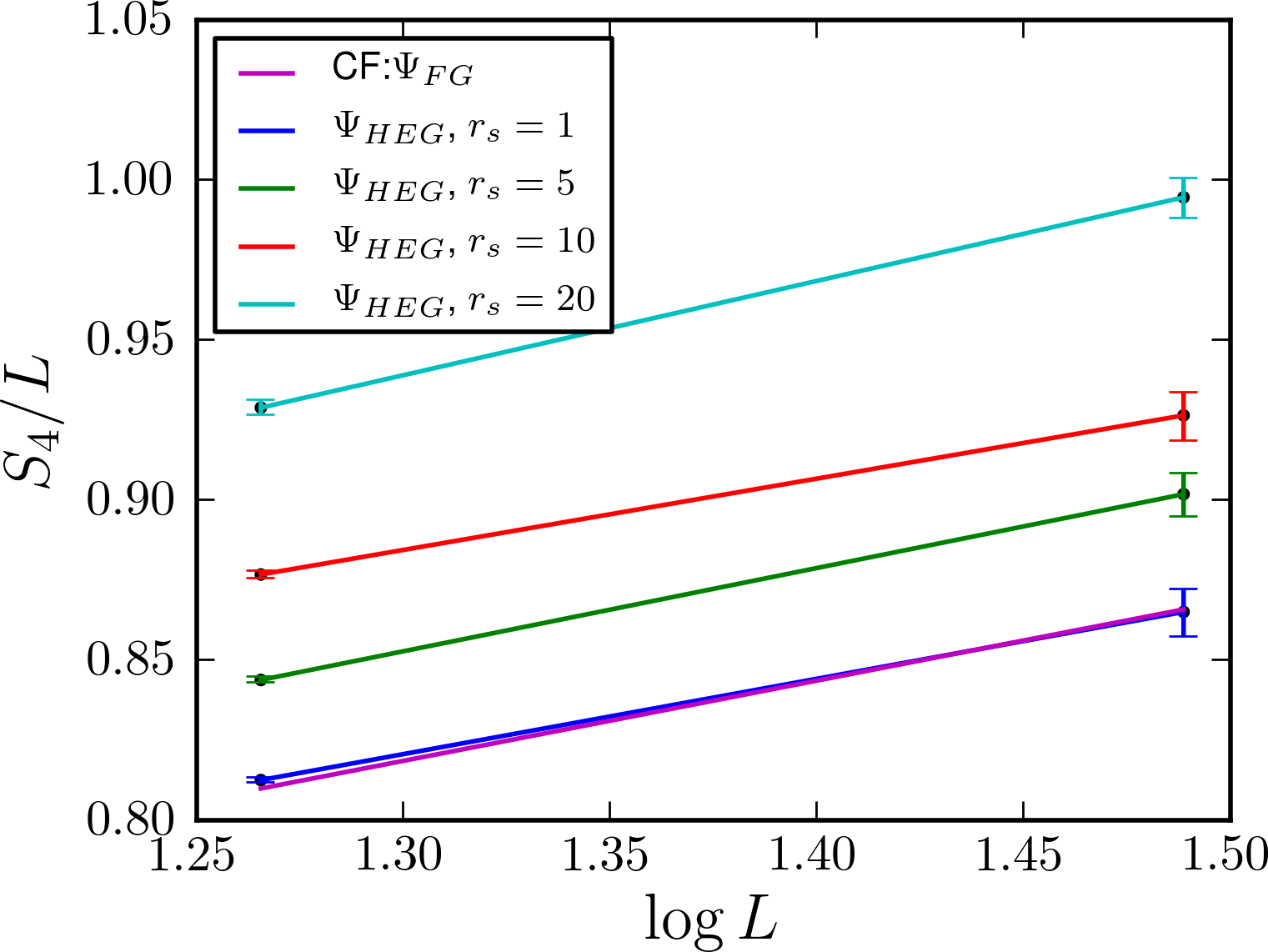}
\caption{ $S_3/L$ and $S_4/L$ plotted against $\log L$. The smallest region contains $\langle N \rangle = 4$ particles and the largest $\langle N \rangle =6.25$.
\label{delta_sn_plot_supp}}
\end{figure}

Because the expectation value of the swap operator can be negative when the value is small and the error is large, we are unable to plot the full range of values for $S_3$ and $S_4$ in Figure \ref{delta_sn_plot_supp}. It can be seen that for the region sizes we are able to compute, the qualitative trends for $S_n(L)$ shown for $S_2$ in the Letter also hold for $S_3$ and $S_4$, $S_n^{HEG}(L)>S_n^{FG}(L)$.


\end{document}